\PassOptionsToPackage{unicode}{hyperref}
\PassOptionsToPackage{hyphens}{url}
\PassOptionsToPackage{dvipsnames,svgnames,x11names}{xcolor}
\documentclass[
]{article}
\usepackage{amsmath,amssymb}
\usepackage{lmodern}
\usepackage{iftex}
\ifPDFTeX
  \usepackage[T1]{fontenc}
  \usepackage[utf8]{inputenc}
  \usepackage{textcomp} 
\else 
  \usepackage{unicode-math}
  \defaultfontfeatures{Scale=MatchLowercase}
  \defaultfontfeatures[\rmfamily]{Ligatures=TeX,Scale=1}
\fi
\IfFileExists{upquote.sty}{\usepackage{upquote}}{}
\IfFileExists{microtype.sty}{
  \usepackage[]{microtype}
  \UseMicrotypeSet[protrusion]{basicmath} 
}{}
\makeatletter
\@ifundefined{KOMAClassName}{
  \IfFileExists{parskip.sty}{%
    \usepackage{parskip}
  }{
    \setlength{\parindent}{0pt}
    \setlength{\parskip}{6pt plus 2pt minus 1pt}}
}{
  \KOMAoptions{parskip=half}}
\makeatother
\usepackage{xcolor}
\usepackage{longtable,booktabs,array}
\usepackage{calc} 
\usepackage{etoolbox}
\makeatletter
\patchcmd\longtable{\par}{\if@noskipsec\mbox{}\fi\par}{}{}
\makeatother
\IfFileExists{footnotehyper.sty}{\usepackage{footnotehyper}}{\usepackage{footnote}}
\makesavenoteenv{longtable}
\usepackage{graphicx}
\makeatletter
\def\maxwidth{\ifdim\Gin@nat@width>\linewidth\linewidth\else\Gin@nat@width\fi}
\def\maxheight{\ifdim\Gin@nat@height>\textheight\textheight\else\Gin@nat@height\fi}
\makeatother
\setkeys{Gin}{width=\maxwidth,height=\maxheight,keepaspectratio}
\makeatletter
\def\fps@figure{htbp}
\makeatother
\NewDocumentCommand\citeproctext{}{}
\NewDocumentCommand\citeproc{mm}{%
  \begingroup\def\citeproctext{#2}\cite{#1}\endgroup}
\makeatletter
 \let\@cite@ofmt\@firstofone
 \def\@biblabel#1{}
 \def\@cite#1#2{{#1\if@tempswa , #2\fi}}
\makeatother
\newlength{\cslhangindent}
\newlength{\csllabelwidth}
\newenvironment{CSLReferences}[2] 
 {\begin{list}{}{%
  \setlength{\itemindent}{0pt}
  \setlength{\leftmargin}{0pt}
  \setlength{\parsep}{0pt}
  \ifodd #1
   \setlength{\leftmargin}{\cslhangindent}
   \setlength{\itemindent}{-1\cslhangindent}
  \fi
  \setlength{\itemsep}{#2\baselineskip}}}
 {\end{list}}
\usepackage{calc}

\ifLuaTeX
\usepackage[bidi=basic]{babel}
\else
\usepackage[bidi=default]{babel}
\fi
\babelprovide[main,import]{american}

\def\languageshorthands#1{}
\ifLuaTeX
  \usepackage{selnolig}  
\fi
\IfFileExists{bookmark.sty}{\usepackage{bookmark}}{\usepackage{hyperref}}
\IfFileExists{xurl.sty}{\usepackage{xurl}}{} 
\hypersetup{
  pdftitle={Networktools: A Header-only C++ Library for Network
Optimization Problems},
  pdfauthor={Morgan Chopin},
  pdflang={en-US},
  colorlinks=true,
  linkcolor={Maroon},
  filecolor={Maroon},
  citecolor={Blue},
  urlcolor={Blue},
  pdfcreator={LaTeX via pandoc}}

\title{Networktools: A Header-only \texttt{C++} Library for Network
Optimization Problems}

\definecolor{c53baa1}{RGB}{83,186,161}
\definecolor{c202826}{RGB}{32,40,38}

\usepackage[affil-it]{authblk}
\usepackage{orcidlink}
\author[1%
  ]{Morgan Chopin%
    \,\orcidlink{0000-0002-9668-1300}\,%
    }

\affil[1]{Orange Research, Châtillon, France%
  }
\date{17 November 2026}

\begin{document}
\maketitle

\section{Summary}\label{summary}

Networktools\footnote{\url{https://gitlab.com/Orange-OpenSource/network-optimization-tools/networktools}}
is an open source, header-only \texttt{C++}20 library for graph-based
modeling and optimization, with dedicated algorithms and data structures
for modeling and solving telecommunications network optimization
problems. More precisely, the library provides general purpose graph
data structures and algorithms for tasks such as efficient graph
traversal, flow computation, matching and connectivity analysis,
together with a generic API to linear and mixed integer linear
programming solvers. Additionally, a dedicated traffic engineering
module provides further specialized data structures and algorithms that
can be used to simulate routing, assess how router or IP/optical link
failures affect traffic loads, and optimize routing configurations such
as link weights and/or segment routing paths.

\section{Statement of Need}\label{statement-of-need}

Operations research applied in the context of telecommunications network
combines general graph problems, such as connectivity analysis and flow
computation, with domain-specific questions such as how traffic is
routed under normal operation and after network failures. General
purpose graph libraries such as LEMON (\citeproc{ref-lemon2011}{Dezső et
al., 2011}) provide reusable algorithms and data structures, while
traffic engineering frameworks such as REPETITA
(\citeproc{ref-REPETITA}{Steven Gay, 2017}) support repeatable
evaluation of traffic engineering algorithms. Combining these
capabilities in a research work can require separate data
representations and integration code which can be time consuming.
Networktools addresses this need by bringing graph algorithms, traffic
engineering simulation and optimization algorithms along with a high
level interface to several LP/MIP solvers together in a single,
header-only \texttt{C++} library.

\section{State of the Field}\label{state-of-the-field}

LEMON (\citeproc{ref-lemon2011}{Dezső et al., 2011}) is a
well-established \texttt{C++} graph-optimization library and the
principal foundation of networktools. Rather than using LEMON as an
external dependency, our library incorporates and extends it by
introducing new features and algorithms, bug fixes, and substantial
performance improvements (see \hyperref[software-design]{Software
Design} and \hyperref[benchmarks]{Benchmarks} sections). Among the
additions, one can find new graph representations such as Compressed
Sparse Row (CSR) based digraph and a compile-time configurable shortest
path engine with dynamic weights update feature
(\citeproc{ref-RamalingamReps96}{Ramalingam \& Reps, 1996}). These
components form the core routines of the different traffic engineering
algorithms of networktools. Notice that while LEMON itself provides
LP/MIP interfaces, networktools uses a separate advanced mathematical
programming module derived from
\href{https://developers.google.com/optimization}{OR-Tools} with
additional features.

Open-source traffic engineering tools such as TOTEM
(\citeproc{ref-TOTEM}{Leduc et al., 2006}) and REPETITA
(\citeproc{ref-REPETITA}{Steven Gay, 2017}) provide a toolbox for
network studies and a framework for repeatable evaluation, respectively.
To our knowledge, few reusable libraries combine graph algorithms,
mathematical programming, and routing simulation for developing
combinatorial optimization methods in traffic engineering. Networktools
addresses this gap by letting researchers implement algorithms and
models and evaluate them using routing and failure-analysis components
within the same \texttt{C++} library.

\section{Software Design}\label{software-design}

Inspired in part by the orthodox \texttt{C++} philosophy, networktools
uses \texttt{C++}20 features selectively: it favors explicit copying,
limits exception-based control flow and the use of STL containers, and
selects performance-sensitive algorithm variants at compile time. Custom
containers provide control over allocation and data layout while
retaining standard interfaces where practical. Networktools requires a
\texttt{C++}20 compiler but no mandatory external library (except for
optional mathematical programming backends that require their respective
solvers).

The library is divided into four modules. Rather than describing each of
them in details or listing every implemented algorithm and feature, we
highlight the principal additions and most relevant features

\textbf{Graphs.} This module includes all the graph representations,
algorithms and utilities for the library. Because shortest path
computation is a frequently used subroutine in simulations of network
routing protocols, particular care has been taken in implementing the
Dijkstra algorithm, which is available in several variants. The default
implementation, \texttt{Dijkstra}, allows features such as traversal
direction and none, single or all predecessors tracking to be selected
at compile time. A variant that supports only 16 bit weights but enables
the use of a SIMD-optimized heap structure (\texttt{QuickHeap}) is also
available. The dynamic variant, \texttt{DynamicDijkstra}, updates
shortest paths after weight changes without recomputing all of them from
scratch (\citeproc{ref-RamalingamReps96}{Ramalingam \& Reps, 1996}).
This module also includes an implementation of the Customized
Contraction Hierarchy (CCH) algorithm that allows to recompute the
shortest paths quickly by exploiting so-called shortcuts even when a
large number of weight changes
(\citeproc{ref-DibbeltStrasserWagner16}{Dibbelt et al., 2016}). This
module provides also several other graph algorithms for tree
decomposition and chordality, multi-commodity flow, inverse shortest
paths, feedback vertex set and betweeness centrality. Graph I/O includes
\href{https://graph-tool.skewed.de/static/docs/stable/gt_format.html}{binary}
and \href{https://github.com/svissicchio/Repetita}{REPETITA} read/write
support, together with readers for formats such as
\href{https://networkx.org/documentation/stable/reference/readwrite/json_graph.html}{NetworkX
JSON}, \href{https://jsongraphformat.info/}{JSON Graph Format}, and
\href{https://sndlib.put.poznan.pl}{SNDlib}, enabling reuse of existing
network datasets.

\textbf{Mathematical programming.} This module provides a unified high
level interface to several LP/MIP solvers (CPLEX, Gurobi, HiGHS, CBC,
GLPK). The primary purpose of utilizing this module is to abstract the
implementation details of the mathematical model from the specific
solver employed for its resolution. This abstraction facilitates
seamless deployment and adaptation of the code to the solver installed
on the platform where the library is deployed. The code of this module
is derived from a significantly modified version of the Google OR-tools'
lp module. The module is header-only, uses networktools containers, and
does not require Abseil or Protocol Buffers. Its additions include
graph-indexed variables, Farkas dual values for infeasibility analysis,
and solver-specific features such as CPLEX cut callbacks and a HiGHS
adapter.

\begin{figure}
\centering
\includegraphics[width=\linewidth,height=3cm,keepaspectratio]{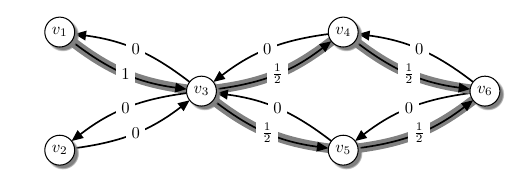}
\caption{Illustration of routing a single unit demand through a
bidirected network from node \(v_1\) to \(v_6\) using the OSPF/IS-IS
routing protocol with Equal-Cost-Multi-Path (ECMP) rule. All arcs have
unit weights and unit capacity. First, the shortests paths from \(v_1\)
to \(v_6\) are computed which are \(v_1,v_3,v_4,v_6\) and
\(v_1,v_3,v_5,v_6\). Second, the demand is routed along these two
shortest paths as follows. Arc \(v_1 v_3\) carries the full demand. At
\(v_3\), the two eligible next-hop arcs, toward \(v_4\) and \(v_5\),
each receive half of the incoming flow following the ECMP rule. Then
each branch forwards its traffic to the final node \(v_6\). The value on
each arc corresponds to its load or saturation (i.e ratio between the
arc's flow and arc's capacity). \label{fig:igp}}
\end{figure}

\textbf{Traffic engineering.} This module provides traffic engineering
optimization and simulation algorithms. These include algorithms for
simulating segment routing (\texttt{SegmentRouting}) and OSPF/IS-IS
routing protocols (\texttt{ShortestPathRouting}), as well as algorithms
and mathematical models for optimizing the routing configurations (see
\autoref{fig:igp}). Dynamic variants of these simulation algorithms were
also implemented. These variants allow to quickly update the routing
paths when link weights change, which is particularly useful for
evaluating what-if scenarios efficiently.

\begin{figure}
\centering
\includegraphics[width=\linewidth,height=6.5cm,keepaspectratio]{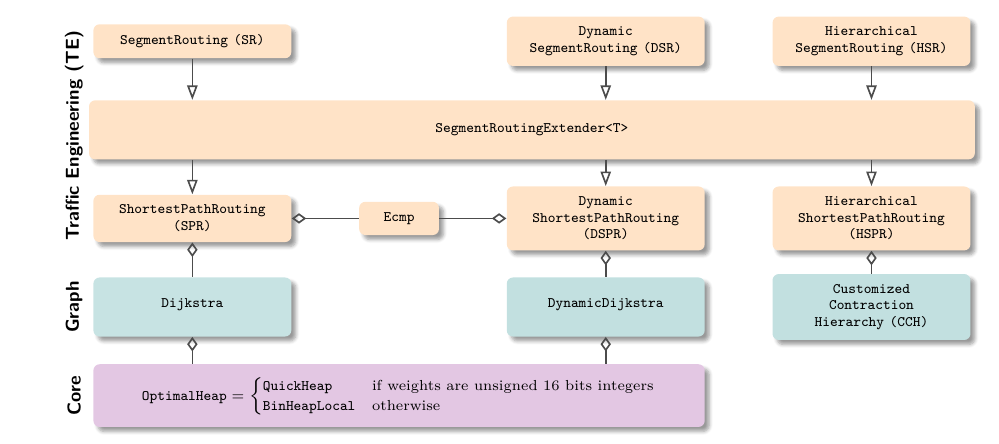}
\caption{Overview of the class hierarchy for shortest path and routing
simulation algorithms. Open triangle arrows denote inheritance while
open diamond arrows denote aggregation.\label{fig:sp-architecture}}
\end{figure}

All of these implementations are based on the Dijkstra implementation
described above and therefore benefit automatically from SIMD
optimization when applicable. Additionally, this module comes with
original simulation algorithms called Hierarchical Shortest Path Routing
(HSPR) and Hierarchical Segment Routing (HSR) that simulates OSPF/IS-IS
routing protocols and segment routing, respectively, using the CCH
algorithm instead of Dijkstra. These implementations are particularly
useful when it comes to evaluate quickly a change of large number of
weights in the network. See \autoref{fig:sp-architecture} for a summary
of the main the routing protocols simulation algorithms and their
relationships. The module also propose several realistic topology and
traffic matrix generators based on state of the art methods
(\citeproc{ref-Nucci2005}{Nucci et al., 2005};
\citeproc{ref-igen}{Quoitin et al., 2009}; \citeproc{ref-inet02}{Winick
\& Jamin, 2002}).

\textbf{Core.} The aforementioned modules rely on the data structures
and common algorithms provided by the core module. Arrays form the
storage foundation for a vast majority of data structures and algorithms
throughout networktools. Their widespread use makes careful
implementation essential for controlling memory allocation, element
lifetimes, and cache locality. The primary and most widely used array
type is \texttt{DynamicArray} supporting configurable growth and
optional small-size optimization. The core module also provides
contiguous multidimensional arrays, bit arrays, linked lists, heaps, and
other utilities. To make potentially costly copies explicit, many owning
types disable implicit copying and instead provide an explicit
duplication operation through \texttt{nt::clone()}.

\section{Benchmarks}\label{benchmarks}

The experiments were performed on a Ubuntu 22.04.1 LTS server equipped
with Intel Xeon Gold 6330 processors and 384 GB RAM. The code was
compiled with GCC 11.4.0 and options
\texttt{-\/-std=c++20\ -O3\ -DNDEBUG\ -march=native\ -DNT\_USE\_SIMD}.
The benchmark instances used for these experiments is composed of 18
network topologies and traffic matrices partitioned into 6 instances
taken from (\citeproc{ref-challenge2026}{Benhamiche et al., 2026})
(ROADEF26), 8 generated with igen tool (\citeproc{ref-igen}{Quoitin et
al., 2009}) (IGEN), and 4 generated with inet tool
(\citeproc{ref-inet02}{Winick \& Jamin, 2002}) (INET). The smallest
instance \texttt{igen20} has 20 nodes, 80 arcs and 222 demands, while
the largest one \texttt{inet6000} has 6000 nodes, 21698 arcs, and 360651
demands. Unless stated otherwise, reported running times are obtained
from the average of 20 repetitions.

\begin{figure}
\centering
\includegraphics[keepaspectratio]{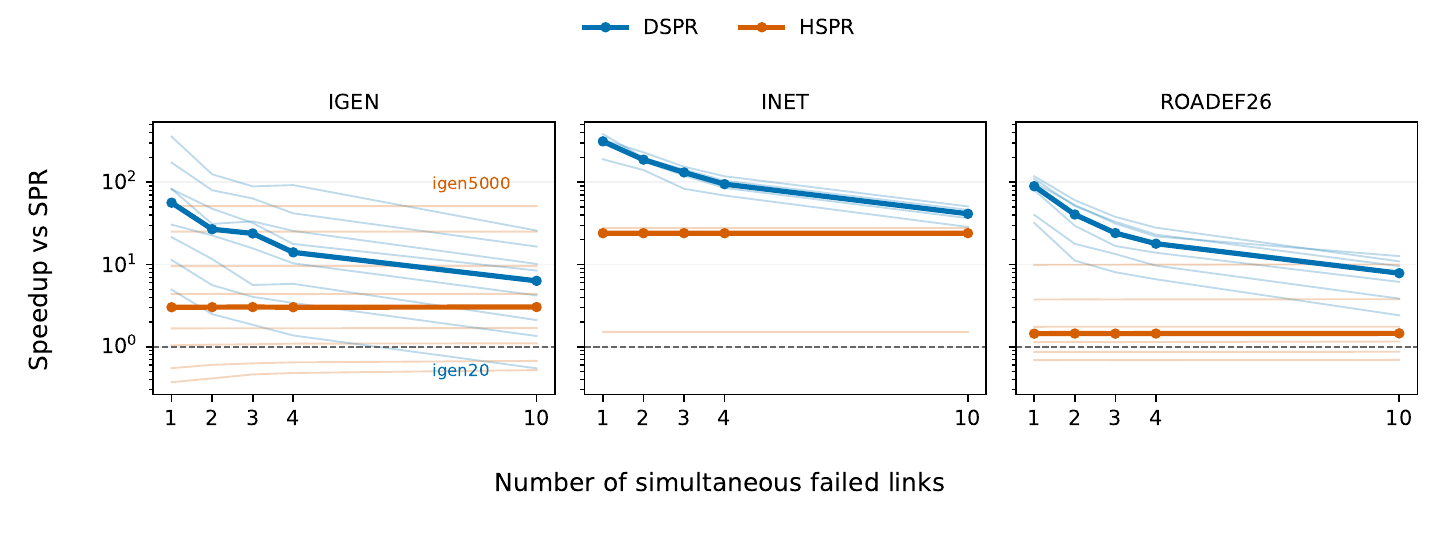}
\caption{Speedup for simulated multi-link failures relative to running a
full recomputation for each scenario using Shortest Path Routing (SPR)
algorithm. More precisely, for each topology and a fix number of
simultaneously failed links, speedup is SPR's mean running time per
scenario divided by that of Dynamic Shortest Path Routing (DSPR) or
Hierarchical Shortest Path Routing (HSPR). Thick lines show the median
speedup across topologies within the same dataset and thin lines show
individual topologies speedup. Values \textgreater{} 1 indicate faster
running time than SPR. HSPR preprocessing time is
excluded.\label{fig:plot-failure}}
\end{figure}

First we consider the problem of computing the induced saturation of a
network in case of multiple link failure. For each topology and each
\(k \in \{1,2,3,4, 10\}\), we instantiate 64 scenarios containing \(k\)
randomly chosen links that fail. We test up to ten simultaneous link
failures, motivated by measurements in an operational IP backbone in
which observed shared optical-failure events affected at most ten IP
links (\citeproc{ref-Markopoulou2008}{Markopoulou et al., 2008}). We
compare DSPR and HSPR with the brute force approach that consists in
calling SPR to recompute all the shortest routing paths for every
scenario. \autoref{fig:plot-failure} shows that DSPR can be more than
two orders of magnitude faster than SPR. The median speedup of DSPR over
SPR decreases from \(56\times\) to \(6.3\times\) on IGEN, from
\(314\times\) to \(41\times\) on INET, and from \(90\times\) to
\(7.8\times\) on ROADEF26 as \(k\) increases from 1 to 10. The median
HSPR speedup remains close to \(3.0\times\), \(24\times\), and
\(1.45\times\), respectively. Thus DSPR remains faster on median even at
\(k=10\), while HSPR's speedup varies little with \(k\). However, HSPR
may become preferable for larger numbers of changed weights, as
previously observed in (\citeproc{ref-DangBCCW23}{Dang et al., 2023}).

\begin{figure}
\centering
\includegraphics[width=\linewidth,height=6.5cm,keepaspectratio]{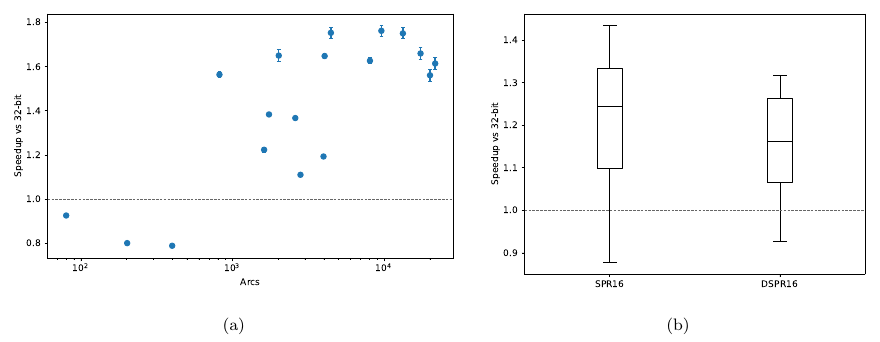}
\caption{Speedup of the 16 bit implementation over the 32 bit
implementation. Values \textgreater{} 1 favor the 16 bit implementation.
(a) Dijkstra16 speedup over Dijkstra32 with respect to the number of
arcs. (b) Boxplot of SPR16 and DSPR16 speedups over SPR32 and DSPR32
across instances, respectively. \label{fig:plot-dijkstra16}}
\end{figure}

In a second series of experiments, we measure the running time
improvement of using the SIMD optimized heap \texttt{QuickHeap} for
Dijkstra when the weights are encoded as unsigned 16 bits integer (see
\autoref{fig:sp-architecture}). Hereafter, we call this variant
Dijkstra16 and the regular version as Dijkstra32. Additionnaly, we study
the improvments of using Dijkstra16 as the shortest path computation
backend for SPR and DSPR. We denote by SPR16/DSPR16 (resp. SPR32/DSPR32)
the variant of SPR/DSPR when Dijkstra16 (resp. Dijkstra32) is used as a
subroutine. Notice that the weights of every instance have been scaled
to fit into 16 bits integers for a fair comparison between the 16 and 32
bits variants. \autoref{fig:plot-dijkstra16}(a) compares the 16 bits and
32 bits Dijkstra implementations. For each instance, we select 16
sources uniformly at random. Then, we measure the total time spent by
Dijkstra to compute the shortest paths from every selected source. The
reported time is the mean over the 20 repetitions of the previous
computation. We remark that Dijkstra16 can be slower on the smallest
networks (namely \texttt{igen20}, \texttt{igen50} and \texttt{igen100}),
whereas on larger networks it is generally faster, reaching
approximately \(1.8\times\) speedup. \autoref{fig:plot-dijkstra16}(b)
shows that Dijkstra16 also improves the running time of SPR and DSPR on
most instances (the values below 1 come from the smallest instances
\texttt{igen20} and \texttt{igen50}). Their median speedups are
\(1.24\times\) and \(1.16\times\), respectively.

The following table reports the mean running times in ms underlying
\autoref{fig:plot-dijkstra16} for four representative instances. Each
speedup in the figure is the corresponding 32-bit mean divided by the
16-bit mean. Notice that DSPR is systematically slower than SPR due to
the additional operations needed to allow dynamic weight updates.
Consequently, it is recommended to use SPR over DSPR when no weight
changes is needed.

\begin{longtable}[]{@{}lrrrrrr@{}}
\toprule\noalign{}
Instance & Dijkstra32 & Dijkstra16 & SPR32 & SPR16 & DSPR32 & DSPR16 \\
\midrule\noalign{}
\endhead
\bottomrule\noalign{}
\endlastfoot
igen50 & 0.014 & 0.0175 & 0.0886 & 0.101 & 0.173 & 0.177 \\
igen1000 & 0.940 & 0.571 & 77.1 & 59.6 & 104 & 84.4 \\
inet6000 & 9.76 & 6.05 & 4520 & 3500 & 5820 & 4670 \\
setB-12 & 1.29 & 0.737 & 94.3 & 70.3 & 126 & 98.6 \\
\end{longtable}

\section{Research Impact Statement}\label{research-impact-statement}

The library has served as the algorithmic backend for several submitted
and peer-reviewed research papers. In (\citeproc{ref-DangBCCW21}{Dang et
al., 2021}, \citeproc{ref-DangBCCW23}{2023}), Monte-Carlo Search and
Nested Rollout Policy Adaptation based algorithms are proposed to solve
the IGP weight-setting problem; the entire experimental part is
implemented using networktools. More recently, networktools has been
used to implement and benchmark the branch-and-cut algorithms for
solving a cyber-security-oriented network optimization problem
(\citeproc{ref-MahjoubDAM}{Chopin et al., 2026}). The library has also
demonstrated its use in domains not directly related to network
optimization (\citeproc{ref-tailhardat2025}{Tailhardat et al., 2025}).
Finally, the library has been used to implement the official checker
program and generate the instances for the EURO/ROADEF 2026 Challenge
(\citeproc{ref-challenge2026}{Benhamiche et al., 2026}). The library is
used to validate and score the solutions submitted by the participating
teams, and the challenge instances are released in formats natively
supported by the library.

\section{AI Usage Disclosure}\label{ai-usage-disclosure}

ChatGPT (OpenAI GPT-6) was used for code completion assistance,
debugging, inline documentation and setting up the benchmarking code.
The AI was also used for paper polishing and grammar corrections. All
AI-assisted outputs were reviewed and validated by the author. The
author developed the paper's scientific content, designed the library,
and implemented its algorithms and data structures, including
substantial adaptations and extensions of code derived from LEMON and
OR-Tools.

\section*{References}\label{references}
\addcontentsline{toc}{section}{References}

\protect\phantomsection\label{refs}

\end{document}